\documentclass[aip,jcp,reprint]{revtex4-1}

\usepackage{amsmath, amsthm, amssymb}
\usepackage{graphicx,color,psfrag}
\usepackage[normalem]{ulem}
\graphicspath{{./Figures/}}

\newcommand{\Tr}[1]{\operatorname{Tr} \left\{ #1 \right\}}

\newcommand{\LU}{{\sc LUMO} }
\newcommand{\HO}{{\sc HOMO} }

\DeclareMathAlphabet{\gcal}{OMS}{cmsy}{m}{n}

\begin{document}

\title{Local entropy of a nonequilibrium fermion system}
\author{Charles\ A.\ Stafford}
\affiliation{Department of Physics, University of Arizona, 1118 East Fourth Street, Tucson, AZ 85721}
\author{Abhay Shastry}
\affiliation{Department of Physics, University of Arizona, 1118 East Fourth Street, Tucson, AZ 85721}
\date{\today}
\begin{abstract}
The local entropy of a nonequilibrium system of independent fermions is investigated, and
analyzed in the context of the laws of thermodynamics.
It is shown that the local temperature and chemical potential can only be expressed in terms of derivatives of the local entropy 
for linear deviations from local equilibrium.  The first law of thermodynamics is shown to lead to an {\em inequality}, not an equality, for
the change in the local entropy as the nonequilibrium state of the system is changed.  
The maximum entropy principle (second law of thermodynamics) is proven:  a nonequilibrium distribution has a
local entropy less than or equal to a local equilibrium distribution satisfying the same constraints.
It is shown that the local entropy of the system tends to zero when the local temperature tends to zero, consistent with the third law of
thermodynamics.
\end{abstract}

\maketitle

\section{Introduction}
\label{sec:intro}

The local entropy of an interacting quantum system is a problem of fundamental interest in many-body
physics,\cite{Zubarev1970,Carneiro1975,Cherng2006,Li2008,Gogolin2016}
quantum field theory,\cite{Kita2006,Kita2010,Prokopec2012,Calabrese2004}
and cosmology.\cite{Bekenstein1973,Bombelli1986,Ashtekar1998,Eling2006,Cai2007} In
this context, the entanglement entropy \cite{Eisert2010} is of central
importance. However, for quantum systems out of equilibrium, even an understanding of the local entropy for
independent particles is lacking.\cite{Polkovnikov2011,Galperin15}
The quest to understand local entropy
of interacting quantum systems without first establishing the results for independent particles may be akin to
seeking a theory of superconductivity without first understanding the noninterating Fermi gas.


In previous work,\cite{Meair14,Stafford16} we showed that local thermodynamic observables such as the temperature and chemical potential can be placed
within the framework of the laws of thermodynamics, even for quantum systems far from equilibrium.  As for the entropy itself, this has so far been shown only
for the third law of thermodynamics, that the local entropy tends to zero as the local temperature tends to zero.\cite{Shastry15}
On the other hand, Esposito, Ochoa, and Galperin have constructed a definition of the local entropy\cite{Galperin15} of a time-dependent resonant level
model that explicitly obeys all laws of thermodynamics even far from equilibrium.  However, their result\cite{Galperin15} does not reduce to the known
result for the entropy in equilibrium.  Moreover, the quantities in their theory cannot be expressed as expectation values of quantum mechanical
operators, calling into question the theoretical basis of their formalism.

In the present article, we propose a definition of the local entropy of a nonequilibrium steady-state
system of independent fermions based entirely on local quantum
observables.  We analyze how this nonequilibrium entropy fits within the framework of the laws of thermodynamics.  We find, contrary to the claims
of Ref.\ \onlinecite{Galperin15}, that the laws of thermodynamics
cannot in general be expressed in differential form in terms of the nonequilibrium entropy.  Rather, such expressions are shown to hold only
to linear order in the deviation from equilibrium, and result in inequalities for systems far from equilibrium, consistent with the maximum entropy principle.
Our detailed analysis of the laws of
thermodynamics in terms of the local nonequilibrium entropy reveals important insights into the statistical mechanics of quantum systems far from
equilibrium.

\section{Entropy Definitions}
\label{sec:definitions}

The starting point for our analysis is the known result for the global entropy of a nonequilibrium system of independent fermions \cite{LandauLifschitz}
\begin{equation}
S= -k_B \sum_{n}[f_{n}\ln{f_{n}}+(1-f_{n})\ln{(1-f_{n})}],
\label{eq:S_LL}
\end{equation}
where $f_n$ is the probability that the $n$th single-particle energy eigenstate (orbital) is occupied.  This result may be derived straightforwardly
from the standard definition
\begin{equation}
S=-k_B \Tr{\hat{\rho} \ln \hat{\rho}} = -k_B \langle \ln\hat\rho\rangle,
\label{eq:S_def}
\end{equation}
where $\hat{\rho}$ is the density matrix of the system.  The density matrix describing a steady state (in or out of equilibrium) is diagonal in the 
energy basis, and for independent fermions may be written as
$\hat{\rho} = \prod_{\otimes n} \hat{\rho}_n$, where the density matrix of a single orbital is
\begin{equation}
\hat{\rho}_n = \left(\begin{array}{cc} f_n & 0 \\ 0 & 1-f_n \end{array}\right).
\end{equation}
Then $\langle \ln\hat{\rho}\rangle = \sum_n \langle \ln\hat{\rho}_n\rangle$, which leads directly to Eq.\ (\ref{eq:S_LL}).

Since we are interested in open quantum systems with (generically) continuous spectra, the sum over states in Eq.\ (\ref{eq:S_LL}) may be replaced
by an energy integral
\begin{eqnarray} 
S[f(\omega)] & = & -k_B \int_{-\infty}^\infty d\omega g(\omega) [f(\omega)\ln f(\omega) \nonumber \\
& & + (1-f(\omega))\ln(1-f(\omega))],
\label{eq:S_global}
\end{eqnarray}
where $g(\omega)\equiv\Tr{A(\omega)}$ is the density of states of the system and 
\begin{equation}
A(\omega) = \frac{1}{2\pi i} \left[ G^<(\omega) - G^>(\omega)\right]
\label{eq:A_def}
\end{equation}
is the spectral function.  
$G^<(\omega)$ and $G^>(\omega)$ are Fourier transforms of the nonequilibrium Green's functions \cite{Stefanucci13}
\begin{eqnarray}
G^<({\bf x},{\bf x'},t) & \equiv & i \langle \hat{\psi}^\dagger({\bf x},0) \hat{\psi}({\bf x}',t)\rangle, \\
G^>({\bf x},{\bf x'},t) & \equiv & -i \langle \hat{\psi}({\bf x}',t) \hat{\psi}^\dagger({\bf x},0) \rangle, 
\end{eqnarray}
where $\hat{\psi}^\dagger({\bf x},t)$ and $\hat{\psi}({\bf x},t)$ are fermion creation and annihilation operators.\footnote{Spin indices have been 
suppressed for simplicity; we restrict consideration to systems without broken spin-rotation invariance in this article.}
The distribution function $f(\omega)$ may be defined in terms of the Green's functions of the quantum system as
\begin{equation}
f(\omega) \equiv \frac{\Tr{G^<(\omega)}}{2\pi i g(\omega)}.
\label{eq:f_def} 
\end{equation}
See Ref.\ \onlinecite{Ness14} for a discussion of nonequilibrium distribution functions.

\subsection{Local entropy}
\label{sec:local_entropy}

In order to define a local entropy for a nonequilibrium quantum system, we consider 
the projection operator
\begin{equation}
\hat{P}({\bf x}) \equiv |{\bf x}\rangle \langle {\bf x}|
\label{eq:Px_def}
\end{equation}
satisfying the completeness relation
\begin{equation}
\int d^3 x \, \hat{P}({\bf x}) =  1.
\end{equation}
The local density of states is then
\begin{equation}
g(\omega;{\bf x}) \equiv \Tr{\hat{P}({\bf x}) A(\omega)} = \langle {\bf x}| A(\omega)|{\bf x}\rangle
\label{eq:gofx_def}
\end{equation}
and the local distribution function is 
\begin{equation}
f(\omega;{\bf x}) \equiv \frac{\Tr{\hat{P}({\bf x}) G^<(\omega)}}{2\pi i \Tr{\hat{P}({\bf x}) A(\omega)}}
= \frac{G^<({\bf x}, {\bf x}, \omega)}{2\pi i g(\omega; {\bf x})}.
\label{eq:fs_def}
\end{equation}
Note that these quantities agree with the definitions \cite{Stafford16} 
of the local spectrum and local distribution function sampled by a probe  for the case of
a broad-band probe coupled locally to the system by a tunneling-width matrix
$\Gamma^p({\bf x})=\gamma_p |{\bf x}\rangle \langle {\bf x}|$.

Our ansatz \cite{Shastry15} for the local entropy of a nonequilibrium system of independent fermions is based on the global nonequilibrium entropy
formula (\ref{eq:S_global}), but formulated in terms of the local observables $g(\omega; {\bf x})$ and $f(\omega;{\bf x})$:
\begin{eqnarray}
S({\bf x}) & \equiv & S[f(\omega;{\bf x})] \nonumber \\
& = & -k_B \int_{-\infty}^{\infty}d\omega g(\omega; {\bf x})[f(\omega; {\bf x})\ln{f(\omega; {\bf x})} \nonumber \\
& & \;\;\;\;\;\;\;\;\;
+(1-f(\omega; {\bf x}))\ln{(1-f(\omega; {\bf x}))}].
\label{eq:S_x_def}
\end{eqnarray}
$S({\bf x})$ so defined is the local entropy per unit volume.\footnote{A different normalization of the local entropy was considered
in Ref.\ \onlinecite{Shastry15}, namely, the entropy per orbital.}

\begin{figure}[htb]
        \includegraphics[width=8.0cm]{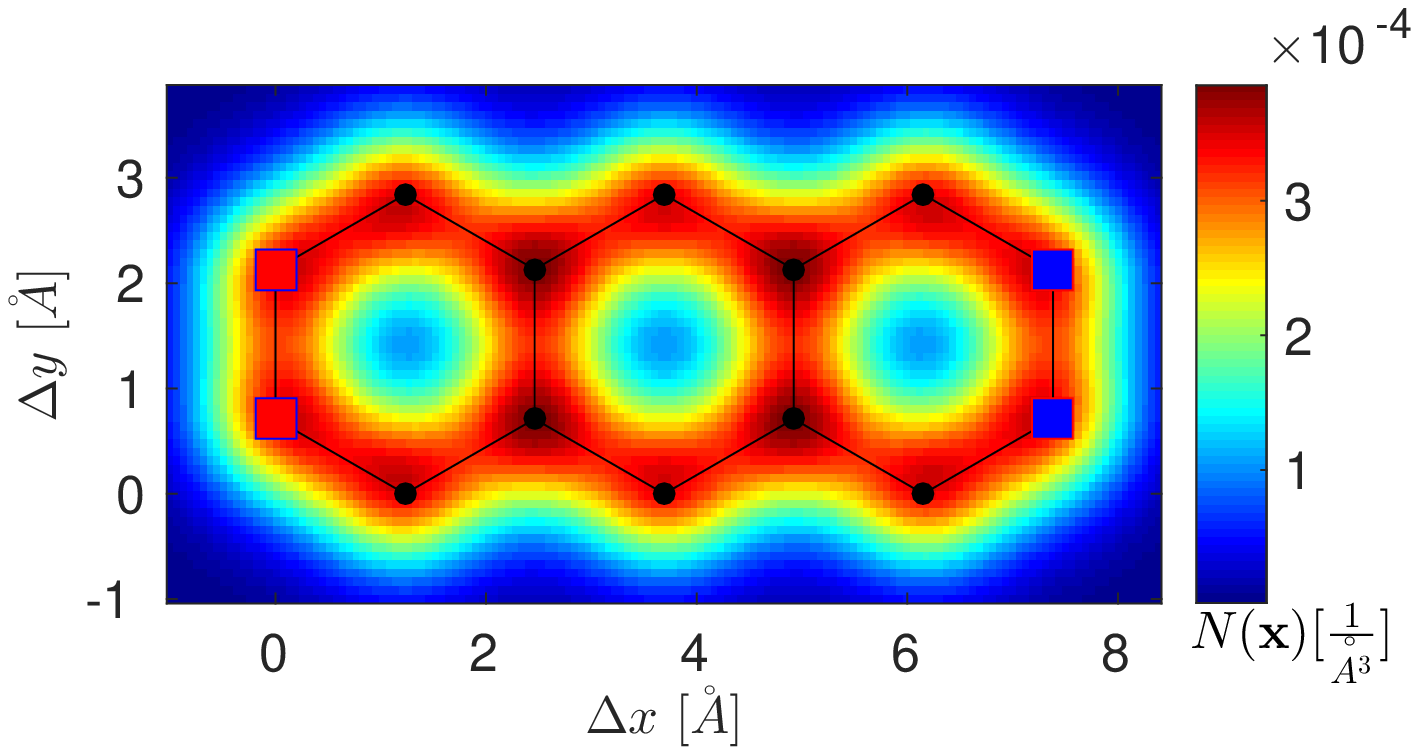}
        \includegraphics[width=8.0cm]{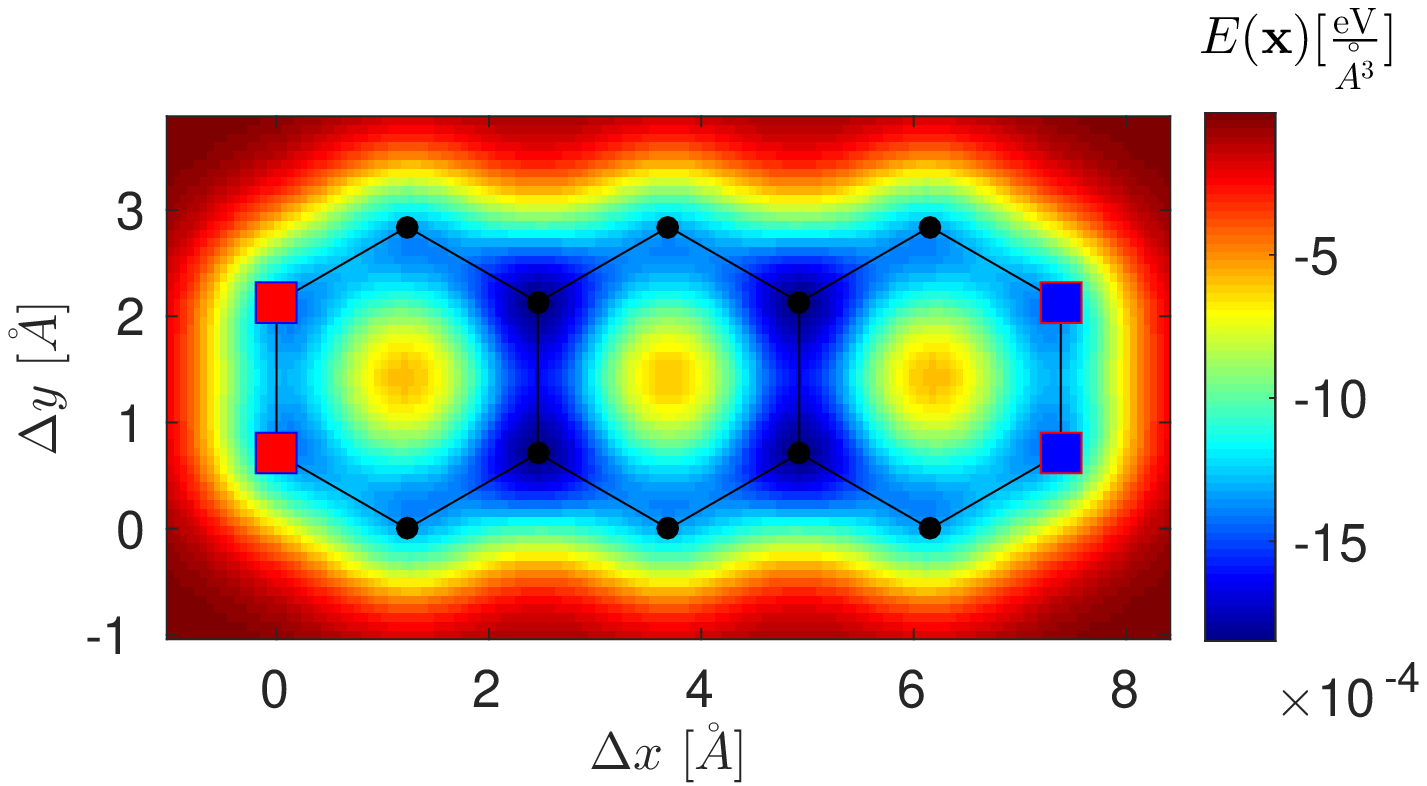}
        \includegraphics[width=8.0cm]{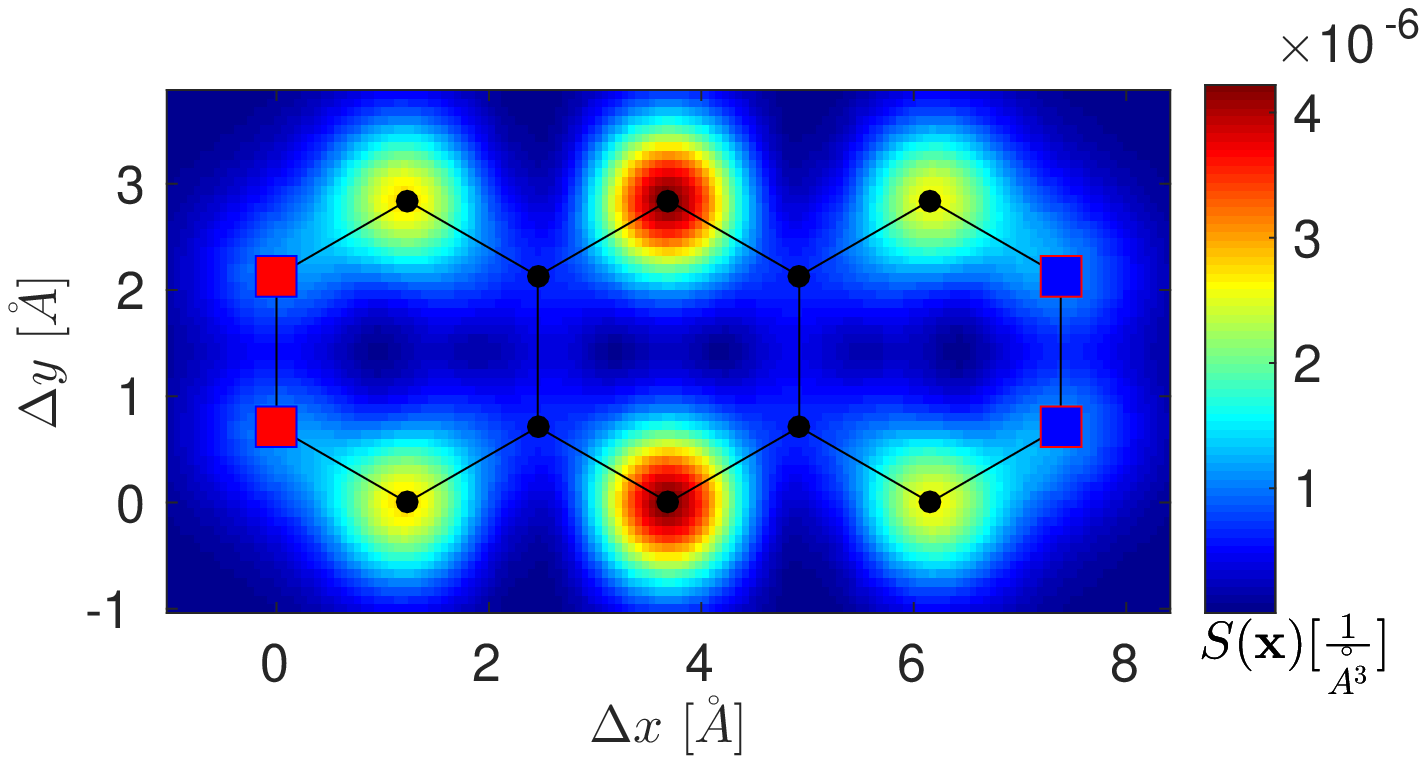}
\caption{
Particle density $N({\bf x})$ (top panel), energy density $E({\bf x})$ (middle panel), and entropy density $S({\bf x})$ (bottom panel) of 
an anthracene molecular junction, evaluated at a height of 2.0\AA\ above the plane of the C nuclei.
Source and drain electrodes held at temperatures of 300K and 100K, respectively, with an electrical bias of 1.5V between them,
are covalently bonded to the molecule at the sites marked by the red and blue squares, respectively.
}
\label{fig:densities}
\end{figure}

The particle density $N({\bf x})$ and energy density $E({\bf x})$ of the nonequilibrium system are
\begin{eqnarray}
N({\bf x}) & = & \int d\omega \, g(\omega; {\bf x}) f(\omega; {\bf x}), \label{eq:Nofx_def}\\
E({\bf x}) & = & \int d\omega \, g(\omega; {\bf x}) \omega f(\omega; {\bf x}).
\label{eq:Eofx_def}
\end{eqnarray}
Fig.\ \ref{fig:densities} shows the nonequilibrium particle, energy, and entropy densities of a single-molecule junction consisting of an anthracene molecule
covalently bonded to source and drain electrodes at the sites marked by the red and blue squares, respectively.  
Source and drain are held at temperatures of 300K and 100K, respectively, with an electrical bias of 1.5V.
$N({\bf x})$ and $E({\bf x})$ have contributions from all of the occupied states of the molecule, while $S({\bf x})$ mainly has contributions from 
electrons in the \LU and 
holes in the \HO of the molecule.

\subsection{Subspace Entropy}
\label{sec:subspace_entropy}

Similarly, the local entropy of a subspace $\alpha$ of a quantum system can be defined with the help of the projection operator
\begin{equation}
\hat{P}_\alpha \equiv \int_{{\bf x}\in\alpha} d^3 x \, \hat{P}({\bf x}).
\label{eq:P_alpha_def}
\end{equation}
The density of states of subspace $\alpha$ is
\begin{equation}
g_\alpha(\omega) \equiv \Tr{\hat{P}_\alpha A(\omega)} = \int_{{\bf x}\in\alpha} d^3 x \,\langle {\bf x}| A(\omega)|{\bf x}\rangle
\label{eq:g_alpha_def}
\end{equation}
and the distribution function of subspace $\alpha$ is
\begin{equation}
f_\alpha(\omega) \equiv \frac{\Tr{\hat{P}_\alpha G^<(\omega)}}{2\pi i \Tr{\hat{P}_\alpha A(\omega)}}
= \frac{\int_{{\bf x}\in\alpha} d^3 x \,G^<({\bf x}, {\bf x}, \omega)}{2\pi i g_\alpha(\omega)}.
\label{eq:f_alpha_def}
\end{equation}
The local entropy of subspace $\alpha$ is
\begin{eqnarray}
S_\alpha & \equiv & S[f_\alpha(\omega)] \nonumber \\
& = & -k_B \int_{-\infty}^{\infty}d\omega g_\alpha(\omega)[f_{\alpha}(\omega)\ln{f_{\alpha}(\omega)} \nonumber \\
& & \;\;\;\;\;\;\;\;\;
+(1-f_{\alpha}(\omega))\ln{(1-f_{\alpha}(\omega))}],
\label{eq:S_alpha_def}
\end{eqnarray}
and is an extensive quantity (not normalized to unit volume).

$S_\alpha$ is not the same as the local entropy defined by the reduced density matrix of the subspace spanned by the projection operator 
$\hat{P}_\alpha$.  Tracing over the rest of the system discards all but the coarsest features of the spectrum if $\hat{P}_\alpha$ is 
highly local, leading to a local entropy formula with little thermodynamic meaning.  Moreover, local properties such as $g(\omega;{\bf x})$
and $f(\omega;{\bf x})$ are clearly measurable by scanning probe techniques and/or near-field photoemission, so it behooves us to seek a local
thermodynamic description of the system in terms of these local observables.

\subsection{Convexity}
\label{sec:convexity}

\begin{figure}[htb]
        \includegraphics[width=7.0cm]{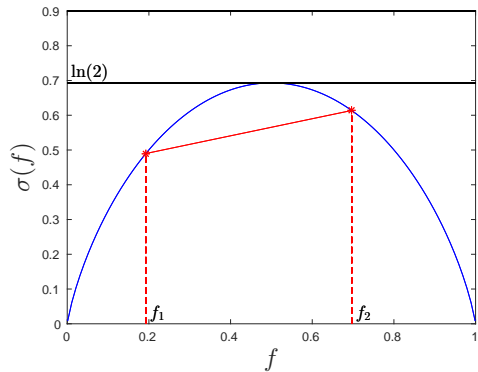}
\caption{
The function $\sigma(f)=-f\ln f -(1-f)\ln(1-f)$, illustrating its convexity:\\
$\sigma(\lambda f_1+(1-\lambda)f_2) \geq \lambda \sigma(f_1) + (1-\lambda)\sigma(f_2)$.
}
\label{fig:convexity_analytic}
\end{figure}

In equilibrium, the distribution function is homogeneous throughout the system, $f(\omega)=f_\alpha(\omega)=f(\omega;{\bf x})$.
This implies that the local entropies are additive in equilibrium:
\begin{equation}
\left.S\right|_{\rm eq} =\sum_\alpha \left.S_\alpha\right|_{\rm eq} = \int d^3 x \left.S({\bf x})\right|_{\rm eq}.
\label{eq:S_additive_eq}
\end{equation}
However, out of equilibrium, the distribution function is in general inhomogeneous.  At each energy, the global distribution function is a weighted 
average of the local distributions:
\begin{equation}
f(\omega) = \frac{\sum_\alpha g_\alpha(\omega) f_\alpha(\omega)}{\sum_\alpha g_\alpha(\omega)}
\end{equation}
and
\begin{equation}
f_\alpha(\omega) = \frac{
\int_{{\bf x}\in\alpha} d^3 x \, g(\omega;{\bf x})f(\omega;{\bf x})
}{
\int_{{\bf x}\in\alpha} d^3 x \, g(\omega;{\bf x})
}.
\end{equation}
The convexity of the function 
$-f\ln f -(1-f)\ln(1-f)$ (see Fig.\ \ref{fig:convexity_analytic}) then implies
\begin{equation}
S \geq \sum_\alpha S_\alpha \geq \int d^3 x S({\bf x}).
\label{eq:S_notadditive_noneq}
\end{equation}
The excess entropy with increasing subsystem size is akin to the {\em entropy of mixing}, since the global distribution function is an energy-dependent 
mixture of the inhomogeneous local distributions.  This effect is to be contrasted with {\em entanglement entropy}, which has the opposite 
sign,\cite{araki1970} and is absent from the present discussion since we consider independent fermions in steady state.

In the following, we focus on an analysis of the local entropy $S({\bf x})$; analogous results for the entropy of an arbitrary subsystem
are given in Appendix \ref{sec:appendix}.

\section{Zeroth Law}
\label{sec:zeroth_law}

Let us consider the conditions under which the local temperature and chemical potential can be uniquely defined in terms of derivatives of the local 
entropy.  The first variation of Eq.\ (\ref{eq:S_x_def}) gives
\begin{equation}
\delta S({\bf x})= k_B \int d\omega \, g(\omega;{\bf x}) \ln\left(\frac{1-f}{f}\right) \delta \! f(\omega;{\bf x}).
\label{eq:deltaS_def}
\end{equation}
Since $0\leq f(\omega;{\bf x}) \leq 1$,\cite{Stafford16} we can write
\begin{equation}
f(\omega;{\bf x}) =\frac{1}{e^{h(\omega;{\bf x})}+1}
\end{equation}
without loss of generality, where $h(\omega;{\bf x}) \in {\cal R}$, so that
$\ln\frac{1-f}{f}=h$.  Then for linear deviations from a local equilibrium distribution
\begin{equation}
f_0(\omega;{\bf x}) = \frac{1}{e^{\beta({\bf x})[\omega-\mu({\bf x})]}+1},
\end{equation}
we have 
\begin{equation}
\delta S({\bf x}) = \frac{1}{T({\bf x})} \left[\delta E({\bf x}) -\mu({\bf x}) \delta N({\bf x})\right],
\label{eq:deltaS_eq}
\end{equation}
where $\beta({\bf x})=\left[k_B T({\bf x})\right]^{-1}$ and 
$N({\bf x})$, $E({\bf x})$ are defined in Eqs.\ (\ref{eq:Nofx_def}), (\ref{eq:Eofx_def}).
Note that Eq.\ (\ref{eq:deltaS_eq}) holds irrespective of the functional form of $\delta \! f(\omega;{\bf x})$, provided $f=f_0$.

We thus have the following definitions of local temperature and chemical potential, valid to linear order in deviations from local equilibrium:
\begin{equation}
\frac{1}{T({\bf x})}  =  \left.\frac{\partial S({\bf x})}{\partial E({\bf x})}\right|_{N({\bf x})} \!\!\!,
\;\;\;\;\;\;
\frac{\mu({\bf x})}{T({\bf x})}  =  -\left.\frac{\partial S({\bf x})}{\partial N({\bf x})}\right|_{E({\bf x})}\!\!\!.
\label{eq:Tandmu_def}
\end{equation}
The ability to express $T$ and $\mu$ in this way for equilibrium systems underlies the universality of equilibrium states codified in the {\em zeroth
law of thermodynamics}.

Far from equilibrium, on the other hand, $h(\omega;{\bf x})\neq \beta({\bf x})[\omega-\mu({\bf x})]$ so that 
$\left.\partial S({\bf x})/\partial E({\bf x})\right|_{N({\bf x})}$ and 
$\left.\partial S({\bf x})/\partial N({\bf x})\right|_{E({\bf x})}$ are not well defined because $\delta S({\bf x})$  [Eq.\ (\ref{eq:deltaS_def})]
depends in detail on the whole function $\delta \! f(\omega;{\bf x})$.
$T({\bf x})$ and $\mu({\bf x})$ can still be uniquely defined \cite{Shastry16} 
far from equilibrium
by an appropriate measurement protocol,\cite{Bergfield2013demon,Meair14,Stafford16}
but they do not have any {\it a priori} relation to variations of the local nonequilibrium entropy.

\section{First Law}
\label{sec:first_law}

\begin{figure}[htb]
        \includegraphics[width=8.5cm]{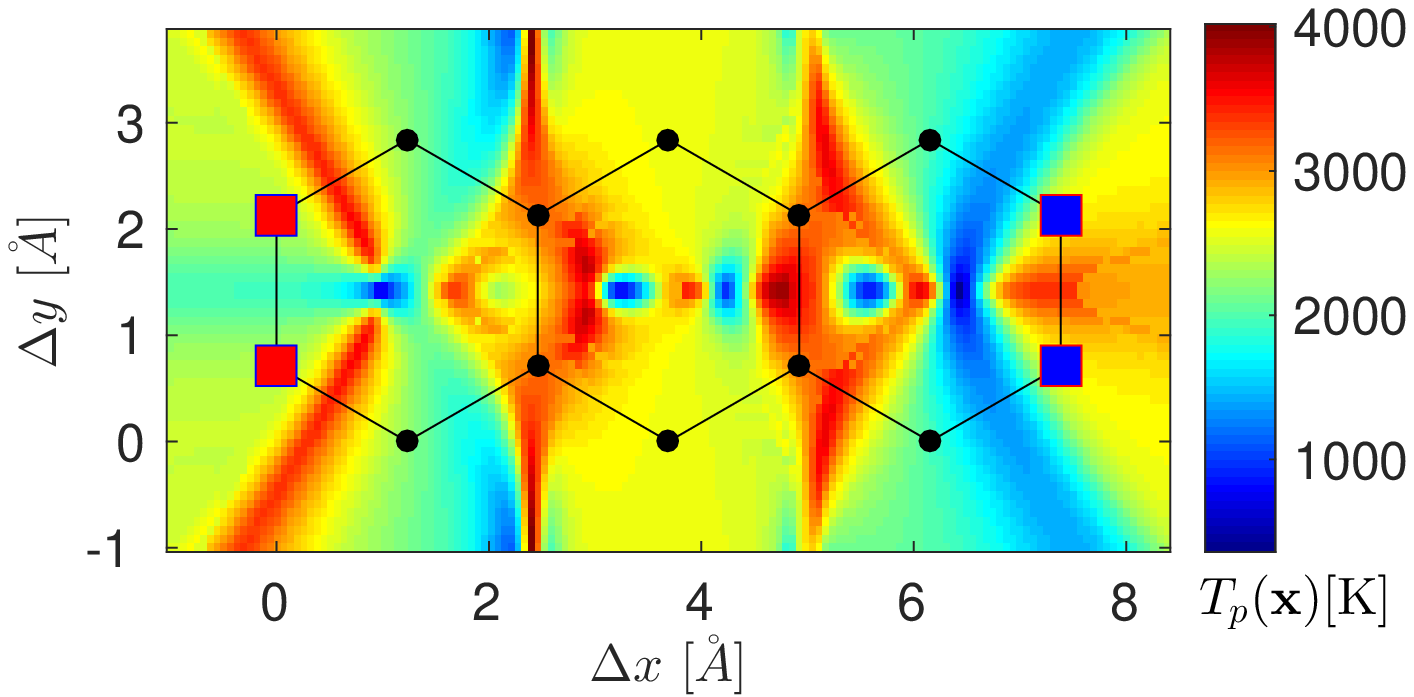}
        \includegraphics[width=8.5cm]{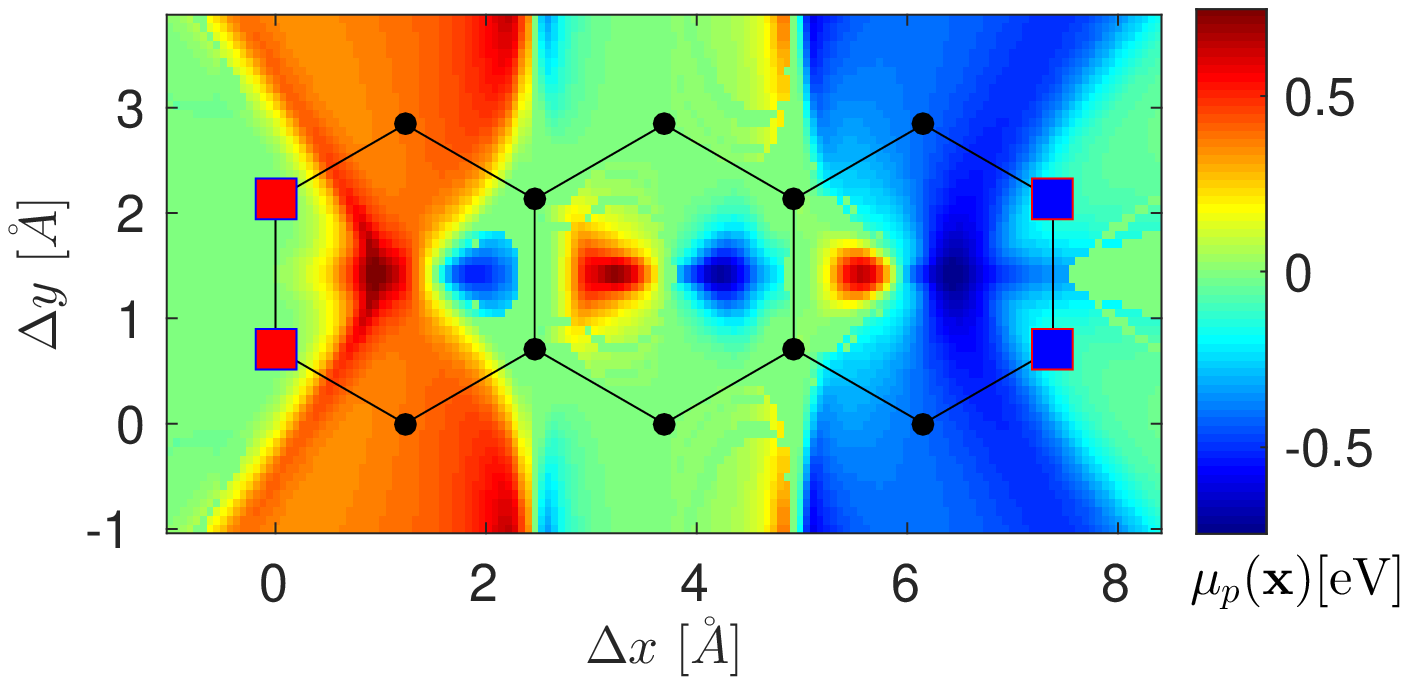}
\caption{
$T_p({\bf x})$ (top panel) and $\mu_p({\bf x})$ (bottom panel) for the same anthracene molecular junction shown in Fig.\ \ref{fig:densities}.
$T_p$ and $\mu_p$ are the temperature and chemical potential 
of a Fermi-Dirac distribution that matches the local particle and
energy densities of the nonequilibrium quantum system, and may be interpreted as the local temperature and 
chemical potential of the system.\cite{Stafford16,Shastry16}
}
\label{fig:Tandmu}
\end{figure}

Eq.\ (\ref{eq:deltaS_eq}) implies that the {\em first law of thermodynamics} governs the change in local entropy for linear deviations from local 
equilibrium.  Let us next consider arbitrarily large deviations from local equilibrium
\begin{equation}
f(\omega, {\bf x}) = f_0(\omega, {\bf x}) + \Delta f(\omega, {\bf x}).
\end{equation}
In order to analyze the change in entropy of the system when it is driven far from a local equilibrium distribution $f_0$, it is useful to
define an auxiliary distribution $f_p(\omega)$, a Fermi-Dirac distribution with temperature $T_p$ and chemical potential $\mu_p$
(see Fig.\ \ref{fig:Tandmu}), that
satisfies the two constraints
\begin{eqnarray}
N({\bf x}) & = & \int d\omega \, g(\omega; {\bf x}) f_p(\omega), \label{eq:Np}\\
E({\bf x}) & = & \int d\omega \, g(\omega; {\bf x}) \omega f_p(\omega),
\label{eq:Ep}
\end{eqnarray}
where $N({\bf x})$ and $E({\bf x})$ are the local particle and energy densities of the nonequilibrium system
defined in Eqs.\ (\ref{eq:Nofx_def}) and (\ref{eq:Eofx_def}), respectively.
That is to say, the local particle density and energy density of the nonequilibrium system with distribution $f(\omega; {\bf x})$ are the same as if the 
local spectrum were populated by the equilibrium distribution $f_p(\omega)$.  $f_p$ is the distribution of a floating broad-band thermoelectric probe coupled
locally to the system.\cite{Stafford16,Shastry16}  

Similarly, we can define \cite{Shastry15} an auxiliary local entropy $S_p({\bf x})$ by replacing $f(\omega; {\bf x})$ by 
$f_p(\omega)$ in Eq.\ (\ref{eq:S_x_def}).  Since $S_p$ is the entropy of an auxiliary equilibrium system, it is a state function obeying the usual
thermodynamic relations.  In particular,
\begin{equation}
S_p({\bf x})-S_0({\bf x}) = \int_{\mu_0,T_0}^{\mu_p,T_p} \frac{dE({\bf x})- \mu({\bf x})dN({\bf x})}{T({\bf x})}.
\end{equation}
In contrast,
the change in $S({\bf x})$ 
for small deviations about a nonequilibrium distribution {\em cannot} be described by Eq.\ (\ref{eq:deltaS_eq}).

\begin{figure}[htb]
        \includegraphics[width=7.0cm]{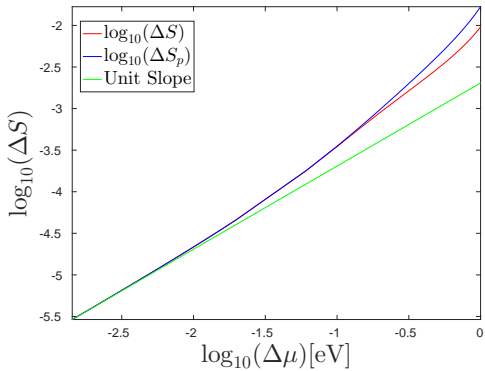}
\caption{
Change $\Delta S$ in the local entropy of site 1 of a two-level quantum system as a function of the electrical bias $\Delta \mu$. 
$\Delta S \simeq \Delta S_p$ 
in the linear-response regime, while $\Delta S \leq \Delta S_p$ in general, where $S_p$ is the entropy of an auxiliary equilibrium
system with the same particle density and energy density as the nonequilibrium system.
}
\label{fig:first_law}
\end{figure}

A Taylor expansion of the integrand in Eq.\ (\ref{eq:S_x_def}) yields
\begin{eqnarray}
S({\bf x})-S_p({\bf x}) & =  & -\frac{k_B}{2}  \int d\omega g(\omega; {\bf x}) \frac{(f-f_p)^2}{f_p(1-f_p)} 
\nonumber \\
& & \mbox{}+ {\cal O}(f-f_p)^3 \leq 0,
\label{eq:second_variation}
\end{eqnarray}
where the inequality is proven below in Sec.\ \ref{sec:second_law}.  
Thus the total change in entropy of the system $\Delta S({\bf x}) = S({\bf x}) - S_0({\bf x})$ cannot be inferred
from the first law of thermodynamics if the final distribution $f(\omega; {\bf x})$ is not an equilibrium distribution.  Instead, the first law gives a
bound on $\Delta S({\bf x})$,
\begin{equation}
\Delta S({\bf x}) \leq \int_{\mu_0,T_0}^{\mu_p,T_p} \frac{dE({\bf x})- \mu({\bf x})dN({\bf x})}{T({\bf x})}.
\end{equation}
This behavior is illustrated in Fig.\ \ref{fig:first_law}, which shows the change in local entropy as a function of electrical bias in a model
two-level quantum system.\footnote{The Hamiltonian of the two-level system has eigenvalues $\pm 1$eV, with couplings $\Gamma_1=\Gamma_2=0.15$eV to
the source and drain reservoirs, subject to a symmetric electrical bias with $\mu_{0}=0$.  Both reservoirs are held at $T=300$K.
}  
$S$ coincides with $S_p$ in the linear-response regime (regime of unit slope on the $\log$-$\log$ plot) but falls below 
$S_p$ for large bias (far from equilibrium).

\section{Maximum Entropy Principle} 
\label{sec:second_law}

\begin{figure}[htb]
        \includegraphics[width=7.0cm]{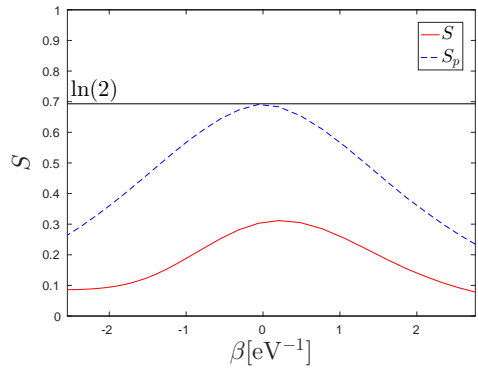}
\caption{
Subsystem entropy $S$ of a two-level quantum system far from equilibrium, plotted versus the inverse temperature $\beta=1/k_B T_p$, as the electrical
bias is varied from 1.6V to 3.2V. 
Values of $\beta <0$ correspond to absolute negative temperatures\cite{Shastry16,Ramsey56} (population inversion).
The figure illustrates the maximum entropy principle, $S\leq S_p$, where $S_p$ is the entropy of an auxiliary equilibrium distribution with
the same local particle and energy densities as the nonequilibrium system.  
}
\label{fig:max_S}
\end{figure}

In this section, we prove the inequality $S({\bf x}) \leq S_p({\bf x})$.  That is to say, the Fermi-Dirac distribution $f_p(\omega)$ is the state
of maximum entropy subject to the constraints (\ref{eq:Nofx_def}) and (\ref{eq:Eofx_def}).  The extremal distribution satisfies 
\begin{eqnarray}
0 & = & \frac{\delta S({\bf x})}{k_B}  + \alpha \delta \! \left(N({\bf x}) - \int d\omega g(\omega;{\bf x}) f(\omega)\right) \nonumber \\
& & \mbox{} + \beta \delta \! \left(E({\bf x}) - \int d\omega g(\omega;{\bf x}) \omega f(\omega)\right),
\label{eq:extremum}
\end{eqnarray}
where $\delta S({\bf x})$ is given by Eq.\ (\ref{eq:deltaS_def}) and $\alpha$ and $\beta$ are Lagrange multipliers.
Eq.\ (\ref{eq:extremum}) may be evaluated straightforwardly, giving
\begin{equation}
0 = \int d\omega g(\omega,{\bf x}) \left[\ln\left(\frac{1}{f}-1\right) -\alpha -\beta \omega\right]\delta\!f(\omega).
\end{equation}
This leads to the maximum entropy distribution
\begin{equation}
f(\omega) = \frac{1}{e^{\beta\omega+\alpha}+1} = f_p(\omega),
\end{equation}
with the usual identification of the Lagrange multipliers 
\begin{equation}
\beta =\frac{1}{k_B T_p}, \;\;\;\;\;\; \alpha = - \frac{\mu_p}{k_B T_p}.
\end{equation}
To verify that this extremum is indeed a maximum, we note that the second variation is negative, as shown in the first line of Eq.\ 
(\ref{eq:second_variation}).
The maximum-entropy principle is illustrated for a model two-level quantum system\cite{Note3} far from equilibrium in Fig.\ \ref{fig:max_S}.

The maximum-entropy principle is a manifestation of the {\em second law of thermodynamics} in a nonequilibrium quantum system: it indicates that the 
system would relax to a local equilibrium distribution of maximum entropy if the forces driving it out of equilibrium were turned off.

\section{Third Law}
\label{sec:third_law}

The local temperature of a quantum system far from equilibrium is thermodynamically meaningful only when
both the local energy and occupation densities are fixed.\cite{Shastry16}  
In particular, a floating broad-band thermoelectric
probe coupled weakly to the system at the point ${\bf x}$ yields the value $T_p$ defined above in Secs.\ \ref{sec:first_law}--\ref{sec:second_law}
(see Refs.\ \onlinecite{Stafford16,Shastry16} for discussion).
Then, one can ask what happens to the local nonequilibrium entropy as the measured value $T_p\rightarrow 0$?

For sufficiently low values of $T_p$, one can evaluate Eq.\ (\ref{eq:S_x_def}) to leading order in the Sommerfeld expansion,
obtaining
\begin{equation}
S({\bf x}) \leq S_p({\bf x}) \sim \frac{\pi^2}{3} g(\mu_p;{\bf x}) k_B^2 T_p \;\; \mbox{as} \;\; T_p\rightarrow 0.
\label{eq:third_law}
\end{equation}
Eq.\ (\ref{eq:third_law}) is a local statement of the {\em third law of thermodynamics} for nonequilibrium fermion systems.  A similar derivation of the
third law using a slightly different definition of local entropy was given in Ref.\ \onlinecite{Shastry15}.

\section{Conclusions}
\label{sec:conclude}

A definition of the local entropy of a nonequilibrium system of independent fermions was proposed, based entirely on local quantum observables.
The laws of thermodynamics were analyzed in terms of differentials of the local nonequilibrium entropy. In general, this procedure only leads to 
equalities for linear deviations from local equilibrium. In certain cases, inequalities were derived for systems far from equilibrium,
consistent with the maximum entropy principle.  Our conclusions also hold for the entropy of an arbitrary subsystem of a nonequilibrium quantum system.

\begin{acknowledgments}
This work was
supported by the U.S.\ Department of Energy
(DOE), Office of Science under Award No.\ DE-SC0006699.
\end{acknowledgments}

\appendix

\section{Analysis of subspace entropy}
\label{sec:appendix}

All of the conclusions concerning the local entropy $S({\bf x})$ of a nonequilibrium fermion system presented in the body of the paper also hold for
the entropy $S_\alpha$ of an arbitrary subspace of the system, defined in Eq.\ (\ref{eq:S_alpha_def}).

In particular, for linear deviations $f_\alpha(\omega)=f_0(\omega) + \delta\! f(\omega)$
from an equilibrium distribution
\begin{equation}
f_0(\omega) = \left[\exp\left(\frac{\omega-\mu_\alpha}{k_B T_\alpha}\right)+1\right]^{-1},
\end{equation}
we have 
\begin{equation}
\delta S_\alpha = \frac{1}{T_\alpha} \left[\delta E_\alpha -\mu_\alpha \delta N_\alpha\right],
\label{eq:deltaS_alpha_eq}
\end{equation}
where 
\begin{eqnarray}
N_\alpha & = & \int d\omega \, g_\alpha(\omega) f_\alpha(\omega), \label{eq:N_alpha_def}\\
E_\alpha & = & \int d\omega \, g_\alpha(\omega) \omega f_\alpha(\omega)
\label{eq:E_alpha_def}
\end{eqnarray}
are the mean number of particles and energy in the subspace, respectively.

The temperature and chemical potential of the subspace are thus given by the following expressions
\begin{equation}
\frac{1}{T_\alpha}  =  \left.\frac{\partial S_\alpha}{\partial E_\alpha}\right|_{N_\alpha} \!\!\!,
\;\;\;\;\;\;
\frac{\mu_\alpha}{T_\alpha}  =  -\left.\frac{\partial S_\alpha}{\partial N_\alpha}\right|_{E_\alpha}\!\!\!,
\label{eq:Tandmu_alpha_def}
\end{equation}
valid to linear order in deviations from local equilibrium.

For large deviations from equilibrium, $f_\alpha(\omega) = f_0(\omega) + \Delta f(\omega)$, the change in subsystem entropy satisfies the
inequality
\begin{equation}
\Delta S_\alpha \leq \int_{\mu_0,T_0}^{\mu_p,T_p} \frac{dE_\alpha- \mu_\alpha dN_\alpha}{T_\alpha},
\end{equation}
where $\mu_p$ and $T_p$ are the chemical potential and temperature of a maximum entropy (Fermi-Dirac) distribution $f_p(\omega)$ satisfying the constraints
\begin{eqnarray}
N_\alpha & = & \int d\omega \, g_\alpha(\omega) f_p(\omega), \label{eq:N_alpha_constraint}\\
E_\alpha & = & \int d\omega \, g_\alpha(\omega) \omega f_p(\omega),
\label{eq:E_alpha_constraint}
\end{eqnarray}
where $N_\alpha$ and $E_\alpha$ are given by Eqs.\ (\ref{eq:N_alpha_def}) and (\ref{eq:E_alpha_def}), respectively.

Finally, for sufficiently low values of $T_p$, one can evaluate Eq.\ (\ref{eq:S_alpha_def}) to leading order in the Sommerfeld expansion,
obtaining \cite{Shastry15}
\begin{equation}
S_\alpha \leq \left.S_\alpha\right|_{f_p} \sim \frac{\pi^2}{3} g_\alpha(\mu_p) k_B^2 T_p \;\; \mbox{as} \;\; T_p\rightarrow 0,
\label{eq:third_law_alpha}
\end{equation}
a statement of the {\em third law of thermodynamics} for a subsystem of a nonequilibrium fermion system.  

\bibliography{refs,RefsLitSurvey}

\begin{thebibliography}{30}%
\makeatletter
\providecommand \@ifxundefined [1]{%
 \@ifx{#1\undefined}
}%
\providecommand \@ifnum [1]{%
 \ifnum #1\expandafter \@firstoftwo
 \else \expandafter \@secondoftwo
 \fi
}%
\providecommand \@ifx [1]{%
 \ifx #1\expandafter \@firstoftwo
 \else \expandafter \@secondoftwo
 \fi
}%
\providecommand \natexlab [1]{#1}%
\providecommand \enquote  [1]{``#1''}%
\providecommand \bibnamefont  [1]{#1}%
\providecommand \bibfnamefont [1]{#1}%
\providecommand \citenamefont [1]{#1}%
\providecommand \href@noop [0]{\@secondoftwo}%
\providecommand \href [0]{\begingroup \@sanitize@url \@href}%
\providecommand \@href[1]{\@@startlink{#1}\@@href}%
\providecommand \@@href[1]{\endgroup#1\@@endlink}%
\providecommand \@sanitize@url [0]{\catcode `\\12\catcode `\$12\catcode
  `\&12\catcode `\#12\catcode `\^12\catcode `\_12\catcode `\%12\relax}%
\providecommand \@@startlink[1]{}%
\providecommand \@@endlink[0]{}%
\providecommand \url  [0]{\begingroup\@sanitize@url \@url }%
\providecommand \@url [1]{\endgroup\@href {#1}{\urlprefix }}%
\providecommand \urlprefix  [0]{URL }%
\providecommand \Eprint [0]{\href }%
\providecommand \doibase [0]{http://dx.doi.org/}%
\providecommand \selectlanguage [0]{\@gobble}%
\providecommand \bibinfo  [0]{\@secondoftwo}%
\providecommand \bibfield  [0]{\@secondoftwo}%
\providecommand \translation [1]{[#1]}%
\providecommand \BibitemOpen [0]{}%
\providecommand \bibitemStop [0]{}%
\providecommand \bibitemNoStop [0]{.\EOS\space}%
\providecommand \EOS [0]{\spacefactor3000\relax}%
\providecommand \BibitemShut  [1]{\csname bibitem#1\endcsname}%
\let\auto@bib@innerbib\@empty
\bibitem [{\citenamefont {Zubarev}\ and\ \citenamefont
  {Kalashnikov}(1970)}]{Zubarev1970}%
  \BibitemOpen
  \bibfield  {author} {\bibinfo {author} {\bibfnamefont {D.}~\bibnamefont
  {Zubarev}}\ and\ \bibinfo {author} {\bibfnamefont {V.}~\bibnamefont
  {Kalashnikov}},\ }\href {\doibase
  http://dx.doi.org/10.1016/0031-8914(70)90143-6} {\bibfield  {journal}
  {\bibinfo  {journal} {Physica}\ }\textbf {\bibinfo {volume} {46}},\ \bibinfo
  {pages} {550 } (\bibinfo {year} {1970})}\BibitemShut {NoStop}%
\bibitem [{\citenamefont {Carneiro}\ and\ \citenamefont
  {Pethick}(1975)}]{Carneiro1975}%
  \BibitemOpen
  \bibfield  {author} {\bibinfo {author} {\bibfnamefont {G.~M.}\ \bibnamefont
  {Carneiro}}\ and\ \bibinfo {author} {\bibfnamefont {C.~J.}\ \bibnamefont
  {Pethick}},\ }\href {\doibase 10.1103/PhysRevB.11.1106} {\bibfield  {journal}
  {\bibinfo  {journal} {Phys. Rev. B}\ }\textbf {\bibinfo {volume} {11}},\
  \bibinfo {pages} {1106} (\bibinfo {year} {1975})}\BibitemShut {NoStop}%
\bibitem [{\citenamefont {Cherng}\ and\ \citenamefont
  {Levitov}(2006)}]{Cherng2006}%
  \BibitemOpen
  \bibfield  {author} {\bibinfo {author} {\bibfnamefont {R.~W.}\ \bibnamefont
  {Cherng}}\ and\ \bibinfo {author} {\bibfnamefont {L.~S.}\ \bibnamefont
  {Levitov}},\ }\href {\doibase 10.1103/PhysRevA.73.043614} {\bibfield
  {journal} {\bibinfo  {journal} {Phys. Rev. A}\ }\textbf {\bibinfo {volume}
  {73}},\ \bibinfo {pages} {043614} (\bibinfo {year} {2006})}\BibitemShut
  {NoStop}%
\bibitem [{\citenamefont {Li}\ and\ \citenamefont {Haldane}(2008)}]{Li2008}%
  \BibitemOpen
  \bibfield  {author} {\bibinfo {author} {\bibfnamefont {H.}~\bibnamefont
  {Li}}\ and\ \bibinfo {author} {\bibfnamefont {F.~D.~M.}\ \bibnamefont
  {Haldane}},\ }\href {\doibase 10.1103/PhysRevLett.101.010504} {\bibfield
  {journal} {\bibinfo  {journal} {Phys. Rev. Lett.}\ }\textbf {\bibinfo
  {volume} {101}},\ \bibinfo {pages} {010504} (\bibinfo {year}
  {2008})}\BibitemShut {NoStop}%
\bibitem [{\citenamefont {Gogolin}\ and\ \citenamefont
  {Eisert}(2016)}]{Gogolin2016}%
  \BibitemOpen
  \bibfield  {author} {\bibinfo {author} {\bibfnamefont {C.}~\bibnamefont
  {Gogolin}}\ and\ \bibinfo {author} {\bibfnamefont {J.}~\bibnamefont
  {Eisert}},\ }\href {http://stacks.iop.org/0034-4885/79/i=5/a=056001}
  {\bibfield  {journal} {\bibinfo  {journal} {Reports on Progress in Physics}\
  }\textbf {\bibinfo {volume} {79}},\ \bibinfo {pages} {056001} (\bibinfo
  {year} {2016})}\BibitemShut {NoStop}%
\bibitem [{\citenamefont {Kita}(2006)}]{Kita2006}%
  \BibitemOpen
  \bibfield  {author} {\bibinfo {author} {\bibfnamefont {T.}~\bibnamefont
  {Kita}},\ }\href {\doibase 10.1143/JPSJ.75.114005} {\bibfield  {journal}
  {\bibinfo  {journal} {Journal of the Physical Society of Japan}\ }\textbf
  {\bibinfo {volume} {75}},\ \bibinfo {pages} {114005} (\bibinfo {year}
  {2006})},\ \Eprint
  {http://arxiv.org/abs/http://dx.doi.org/10.1143/JPSJ.75.114005}
  {http://dx.doi.org/10.1143/JPSJ.75.114005} \BibitemShut {NoStop}%
\bibitem [{\citenamefont {Kita}(2010)}]{Kita2010}%
  \BibitemOpen
  \bibfield  {author} {\bibinfo {author} {\bibfnamefont {T.}~\bibnamefont
  {Kita}},\ }\href {\doibase 10.1143/PTP.123.581} {\bibfield  {journal}
  {\bibinfo  {journal} {Progress of Theoretical Physics}\ }\textbf {\bibinfo
  {volume} {123}},\ \bibinfo {pages} {581} (\bibinfo {year} {2010})},\ \Eprint
  {http://arxiv.org/abs/http://ptp.oxfordjournals.org/content/123/4/581.full.pdf+html}
  {http://ptp.oxfordjournals.org/content/123/4/581.full.pdf+html} \BibitemShut
  {NoStop}%
\bibitem [{\citenamefont {Prokopec}, \citenamefont {Schmidt},\ and\
  \citenamefont {Weenink}(2012)}]{Prokopec2012}%
  \BibitemOpen
  \bibfield  {author} {\bibinfo {author} {\bibfnamefont {T.}~\bibnamefont
  {Prokopec}}, \bibinfo {author} {\bibfnamefont {M.~G.}\ \bibnamefont
  {Schmidt}}, \ and\ \bibinfo {author} {\bibfnamefont {J.}~\bibnamefont
  {Weenink}},\ }\href {\doibase http://dx.doi.org/10.1016/j.aop.2012.09.003}
  {\bibfield  {journal} {\bibinfo  {journal} {Annals of Physics}\ }\textbf
  {\bibinfo {volume} {327}},\ \bibinfo {pages} {3138 } (\bibinfo {year}
  {2012})}\BibitemShut {NoStop}%
\bibitem [{\citenamefont {Calabrese}\ and\ \citenamefont
  {Cardy}(2004)}]{Calabrese2004}%
  \BibitemOpen
  \bibfield  {author} {\bibinfo {author} {\bibfnamefont {P.}~\bibnamefont
  {Calabrese}}\ and\ \bibinfo {author} {\bibfnamefont {J.}~\bibnamefont
  {Cardy}},\ }\href {http://stacks.iop.org/1742-5468/2004/i=06/a=P06002}
  {\bibfield  {journal} {\bibinfo  {journal} {Journal of Statistical Mechanics:
  Theory and Experiment}\ }\textbf {\bibinfo {volume} {2004}},\ \bibinfo
  {pages} {P06002} (\bibinfo {year} {2004})}\BibitemShut {NoStop}%
\bibitem [{\citenamefont {Bekenstein}(1973)}]{Bekenstein1973}%
  \BibitemOpen
  \bibfield  {author} {\bibinfo {author} {\bibfnamefont {J.~D.}\ \bibnamefont
  {Bekenstein}},\ }\href {\doibase 10.1103/PhysRevD.7.2333} {\bibfield
  {journal} {\bibinfo  {journal} {Phys. Rev. D}\ }\textbf {\bibinfo {volume}
  {7}},\ \bibinfo {pages} {2333} (\bibinfo {year} {1973})}\BibitemShut
  {NoStop}%
\bibitem [{\citenamefont {Bombelli}\ \emph {et~al.}(1986)\citenamefont
  {Bombelli}, \citenamefont {Koul}, \citenamefont {Lee},\ and\ \citenamefont
  {Sorkin}}]{Bombelli1986}%
  \BibitemOpen
  \bibfield  {author} {\bibinfo {author} {\bibfnamefont {L.}~\bibnamefont
  {Bombelli}}, \bibinfo {author} {\bibfnamefont {R.~K.}\ \bibnamefont {Koul}},
  \bibinfo {author} {\bibfnamefont {J.}~\bibnamefont {Lee}}, \ and\ \bibinfo
  {author} {\bibfnamefont {R.~D.}\ \bibnamefont {Sorkin}},\ }\href {\doibase
  10.1103/PhysRevD.34.373} {\bibfield  {journal} {\bibinfo  {journal} {Phys.
  Rev. D}\ }\textbf {\bibinfo {volume} {34}},\ \bibinfo {pages} {373} (\bibinfo
  {year} {1986})}\BibitemShut {NoStop}%
\bibitem [{\citenamefont {Ashtekar}\ \emph {et~al.}(1998)\citenamefont
  {Ashtekar}, \citenamefont {Baez}, \citenamefont {Corichi},\ and\
  \citenamefont {Krasnov}}]{Ashtekar1998}%
  \BibitemOpen
  \bibfield  {author} {\bibinfo {author} {\bibfnamefont {A.}~\bibnamefont
  {Ashtekar}}, \bibinfo {author} {\bibfnamefont {J.}~\bibnamefont {Baez}},
  \bibinfo {author} {\bibfnamefont {A.}~\bibnamefont {Corichi}}, \ and\
  \bibinfo {author} {\bibfnamefont {K.}~\bibnamefont {Krasnov}},\ }\href
  {\doibase 10.1103/PhysRevLett.80.904} {\bibfield  {journal} {\bibinfo
  {journal} {Phys. Rev. Lett.}\ }\textbf {\bibinfo {volume} {80}},\ \bibinfo
  {pages} {904} (\bibinfo {year} {1998})}\BibitemShut {NoStop}%
\bibitem [{\citenamefont {Eling}, \citenamefont {Guedens},\ and\ \citenamefont
  {Jacobson}(2006)}]{Eling2006}%
  \BibitemOpen
  \bibfield  {author} {\bibinfo {author} {\bibfnamefont {C.}~\bibnamefont
  {Eling}}, \bibinfo {author} {\bibfnamefont {R.}~\bibnamefont {Guedens}}, \
  and\ \bibinfo {author} {\bibfnamefont {T.}~\bibnamefont {Jacobson}},\ }\href
  {\doibase 10.1103/PhysRevLett.96.121301} {\bibfield  {journal} {\bibinfo
  {journal} {Phys. Rev. Lett.}\ }\textbf {\bibinfo {volume} {96}},\ \bibinfo
  {pages} {121301} (\bibinfo {year} {2006})}\BibitemShut {NoStop}%
\bibitem [{\citenamefont {Cai}\ and\ \citenamefont {Cao}(2007)}]{Cai2007}%
  \BibitemOpen
  \bibfield  {author} {\bibinfo {author} {\bibfnamefont {R.-G.}\ \bibnamefont
  {Cai}}\ and\ \bibinfo {author} {\bibfnamefont {L.-M.}\ \bibnamefont {Cao}},\
  }\href {\doibase 10.1103/PhysRevD.75.064008} {\bibfield  {journal} {\bibinfo
  {journal} {Phys. Rev. D}\ }\textbf {\bibinfo {volume} {75}},\ \bibinfo
  {pages} {064008} (\bibinfo {year} {2007})}\BibitemShut {NoStop}%
\bibitem [{\citenamefont {Eisert}, \citenamefont {Cramer},\ and\ \citenamefont
  {Plenio}(2010)}]{Eisert2010}%
  \BibitemOpen
  \bibfield  {author} {\bibinfo {author} {\bibfnamefont {J.}~\bibnamefont
  {Eisert}}, \bibinfo {author} {\bibfnamefont {M.}~\bibnamefont {Cramer}}, \
  and\ \bibinfo {author} {\bibfnamefont {M.~B.}\ \bibnamefont {Plenio}},\
  }\href {\doibase 10.1103/RevModPhys.82.277} {\bibfield  {journal} {\bibinfo
  {journal} {Rev. Mod. Phys.}\ }\textbf {\bibinfo {volume} {82}},\ \bibinfo
  {pages} {277} (\bibinfo {year} {2010})}\BibitemShut {NoStop}%
\bibitem [{\citenamefont {Polkovnikov}(2011)}]{Polkovnikov2011}%
  \BibitemOpen
  \bibfield  {author} {\bibinfo {author} {\bibfnamefont {A.}~\bibnamefont
  {Polkovnikov}},\ }\href {\doibase
  http://dx.doi.org/10.1016/j.aop.2010.08.004} {\bibfield  {journal} {\bibinfo
  {journal} {Annals of Physics}\ }\textbf {\bibinfo {volume} {326}},\ \bibinfo
  {pages} {486 } (\bibinfo {year} {2011})}\BibitemShut {NoStop}%
\bibitem [{\citenamefont {Esposito}, \citenamefont {Ochoa},\ and\ \citenamefont
  {Galperin}(2015)}]{Galperin15}%
  \BibitemOpen
  \bibfield  {author} {\bibinfo {author} {\bibfnamefont {M.}~\bibnamefont
  {Esposito}}, \bibinfo {author} {\bibfnamefont {M.~A.}\ \bibnamefont {Ochoa}},
  \ and\ \bibinfo {author} {\bibfnamefont {M.}~\bibnamefont {Galperin}},\
  }\href {\doibase 10.1103/PhysRevLett.114.080602} {\bibfield  {journal}
  {\bibinfo  {journal} {Phys. Rev. Lett.}\ }\textbf {\bibinfo {volume} {114}},\
  \bibinfo {pages} {080602} (\bibinfo {year} {2015})}\BibitemShut {NoStop}%
\bibitem [{\citenamefont {Meair}\ \emph {et~al.}(2014)\citenamefont {Meair},
  \citenamefont {Bergfield}, \citenamefont {Stafford},\ and\ \citenamefont
  {Jacquod}}]{Meair14}%
  \BibitemOpen
  \bibfield  {author} {\bibinfo {author} {\bibfnamefont {J.}~\bibnamefont
  {Meair}}, \bibinfo {author} {\bibfnamefont {J.~P.}\ \bibnamefont
  {Bergfield}}, \bibinfo {author} {\bibfnamefont {C.~A.}\ \bibnamefont
  {Stafford}}, \ and\ \bibinfo {author} {\bibfnamefont {P.}~\bibnamefont
  {Jacquod}},\ }\href {\doibase 10.1103/PhysRevB.90.035407} {\bibfield
  {journal} {\bibinfo  {journal} {Phys. Rev. B}\ }\textbf {\bibinfo {volume}
  {90}},\ \bibinfo {pages} {035407} (\bibinfo {year} {2014})}\BibitemShut
  {NoStop}%
\bibitem [{\citenamefont {Stafford}(2016)}]{Stafford16}%
  \BibitemOpen
  \bibfield  {author} {\bibinfo {author} {\bibfnamefont {C.~A.}\ \bibnamefont
  {Stafford}},\ }\href {\doibase 10.1103/PhysRevB.93.245403} {\bibfield
  {journal} {\bibinfo  {journal} {Phys. Rev. B}\ }\textbf {\bibinfo {volume}
  {93}},\ \bibinfo {pages} {245403} (\bibinfo {year} {2016})}\BibitemShut
  {NoStop}%
\bibitem [{\citenamefont {Shastry}\ and\ \citenamefont
  {Stafford}(2015)}]{Shastry15}%
  \BibitemOpen
  \bibfield  {author} {\bibinfo {author} {\bibfnamefont {A.}~\bibnamefont
  {Shastry}}\ and\ \bibinfo {author} {\bibfnamefont {C.~A.}\ \bibnamefont
  {Stafford}},\ }\href {\doibase 10.1103/PhysRevB.92.245417} {\bibfield
  {journal} {\bibinfo  {journal} {Phys. Rev. B}\ }\textbf {\bibinfo {volume}
  {92}},\ \bibinfo {pages} {245417} (\bibinfo {year} {2015})}\BibitemShut
  {NoStop}%
\bibitem [{\citenamefont {Landau}\ and\ \citenamefont
  {Lifshitz}(1980)}]{LandauLifschitz}%
  \BibitemOpen
  \bibfield  {author} {\bibinfo {author} {\bibfnamefont {L.~D.}\ \bibnamefont
  {Landau}}\ and\ \bibinfo {author} {\bibfnamefont {E.~M.}\ \bibnamefont
  {Lifshitz}},\ }\enquote {\bibinfo {title} {Statistical physics},}\ \
  (\bibinfo  {publisher} {Butterworth-Heinemann},\ \bibinfo {year} {1980})\
  pp.\ \bibinfo {pages} {160--161},\ \bibinfo {edition} {3rd}\ ed.\BibitemShut
  {Stop}%
\bibitem [{\citenamefont {Stefanucci}\ and\ \citenamefont {van
  Leeuwen}(2013)}]{Stefanucci13}%
  \BibitemOpen
  \bibfield  {author} {\bibinfo {author} {\bibfnamefont {G.}~\bibnamefont
  {Stefanucci}}\ and\ \bibinfo {author} {\bibfnamefont {R.}~\bibnamefont {van
  Leeuwen}},\ }\href@noop {} {\emph {\bibinfo {title} {Nonequilibrium Many-Body
  Theory Of Quantum Systems: A Modern Introduction}}}\ (\bibinfo  {publisher}
  {Cambridge University Press},\ \bibinfo {year} {2013})\BibitemShut {NoStop}%
\bibitem [{Note1()}]{Note1}%
  \BibitemOpen
  \bibinfo {note} {Spin indices have been suppressed for simplicity; we
  restrict consideration to systems without broken spin-rotation invariance in
  this article.}\BibitemShut {Stop}%
\bibitem [{\citenamefont {Ness}(2014)}]{Ness14}%
  \BibitemOpen
  \bibfield  {author} {\bibinfo {author} {\bibfnamefont {H.}~\bibnamefont
  {Ness}},\ }\href {\doibase 10.1103/PhysRevB.89.045409} {\bibfield  {journal}
  {\bibinfo  {journal} {Phys. Rev. B}\ }\textbf {\bibinfo {volume} {89}},\
  \bibinfo {pages} {045409} (\bibinfo {year} {2014})}\BibitemShut {NoStop}%
\bibitem [{Note2()}]{Note2}%
  \BibitemOpen
  \bibinfo {note} {A different normalization of the local entropy was
  considered in Ref.\ \protect \rev@citealpnum {Shastry15}, namely, the entropy
  per orbital.}\BibitemShut {Stop}%
\bibitem [{\citenamefont {Araki}\ and\ \citenamefont {Lieb}(1970)}]{araki1970}%
  \BibitemOpen
  \bibfield  {author} {\bibinfo {author} {\bibfnamefont {H.}~\bibnamefont
  {Araki}}\ and\ \bibinfo {author} {\bibfnamefont {E.~H.}\ \bibnamefont
  {Lieb}},\ }\href {http://projecteuclid.org/euclid.cmp/1103842506} {\bibfield
  {journal} {\bibinfo  {journal} {Comm. Math. Phys.}\ }\textbf {\bibinfo
  {volume} {18}},\ \bibinfo {pages} {160} (\bibinfo {year} {1970})}\BibitemShut
  {NoStop}%
\bibitem [{\citenamefont {Shastry}\ and\ \citenamefont
  {Stafford}(2016)}]{Shastry16}%
  \BibitemOpen
  \bibfield  {author} {\bibinfo {author} {\bibfnamefont {A.}~\bibnamefont
  {Shastry}}\ and\ \bibinfo {author} {\bibfnamefont {C.~A.}\ \bibnamefont
  {Stafford}},\ }\href {\doibase 10.1103/PhysRevB.94.155433} {\bibfield
  {journal} {\bibinfo  {journal} {Phys. Rev. B}\ }\textbf {\bibinfo {volume}
  {94}},\ \bibinfo {pages} {155433} (\bibinfo {year} {2016})}\BibitemShut
  {NoStop}%
\bibitem [{\citenamefont {Bergfield}\ \emph {et~al.}(2013)\citenamefont
  {Bergfield}, \citenamefont {Story}, \citenamefont {Stafford},\ and\
  \citenamefont {Stafford}}]{Bergfield2013demon}%
  \BibitemOpen
  \bibfield  {author} {\bibinfo {author} {\bibfnamefont {J.~P.}\ \bibnamefont
  {Bergfield}}, \bibinfo {author} {\bibfnamefont {S.~M.}\ \bibnamefont
  {Story}}, \bibinfo {author} {\bibfnamefont {R.~C.}\ \bibnamefont {Stafford}},
  \ and\ \bibinfo {author} {\bibfnamefont {C.~A.}\ \bibnamefont {Stafford}},\
  }\href {\doibase 10.1021/nn401027u} {\bibfield  {journal} {\bibinfo
  {journal} {ACS Nano}\ }\textbf {\bibinfo {volume} {7}},\ \bibinfo {pages}
  {4429} (\bibinfo {year} {2013})}\BibitemShut {NoStop}%
\bibitem [{Note3()}]{Note3}%
  \BibitemOpen
  \bibinfo {note} {The Hamiltonian of the two-level system has eigenvalues $\pm
  1$eV, with couplings $\Gamma _1=\Gamma _2=0.15$eV to the source and drain
  reservoirs, subject to a symmetric electrical bias with $\mu _{0}=0$. Both
  reservoirs are held at $T=300$K.}\BibitemShut {Stop}%
\bibitem [{\citenamefont {Ramsey}(1956)}]{Ramsey56}%
  \BibitemOpen
  \bibfield  {author} {\bibinfo {author} {\bibfnamefont {N.~F.}\ \bibnamefont
  {Ramsey}},\ }\href {\doibase 10.1103/PhysRev.103.20} {\bibfield  {journal}
  {\bibinfo  {journal} {Phys. Rev.}\ }\textbf {\bibinfo {volume} {103}},\
  \bibinfo {pages} {20} (\bibinfo {year} {1956})}\BibitemShut {NoStop}%
\end{thebibliography}%

\end{document}